# Extending LLM-based support for software engineers with ADHD

Aarsh Shah
*University of Calgary*
Calgary, Canada
aarsh.shah@ucalgary.ca

Ronnie de Souza Santos
*University of Calgary*
Calgary, Canada
ronnie.desouzasantos@ucalgary.ca

Italo Santos
*University of Hawaii at Manoa*
United States
isantos3@hawaii.edu

Cleyton Magalhães
*UFRPE*
Brazil
cleyton.vanut@ufrpe.br

Kiev Gama
*Federal University of Pernambuco*
Brazil
kiev@cin.ufpe.br

***Abstract*—Software engineering workflows are often not designed to accommodate the needs of developers with Attention Deficit Hyperactivity Disorder (ADHD), despite known challenges related to task initiation, sustained attention, and completion. At the same time, large language models (LLMs) are increasingly integrated into programming tools, but existing systems do not account for neurodiversity or support structured progression across development tasks. In this work, we present *Tether 2.0*, an LLM based assistant designed to support software engineers with ADHD through workflow oriented interaction across planning, coding, debugging, and review. The tool was named *Tether 2.0* because it builds upon the open source code and foundations of the original *Tether* system. Our approach combines structured interaction modes, activity aware context, and persistent memory to support task progression and continuity. We evaluate the system through expert feedback and hands on use with a software engineer with ADHD. Our results indicate that *Tether 2.0* supports requirement clarification, task decomposition, incremental implementation, debugging, and completion, while enabling users to maintain progress and resume work after interruptions. These findings suggest that LLM based assistants can support direction, sustained progress, and learning in software development when designed with workflow structure and context aware interaction.**

## I. INTRODUCTION

Software engineering produces systems for a diverse society, yet software engineering practices, tools, and environments are often not designed to accommodate different ways of working [1]–[3]. Among the groups that remain underrepresented in both research and practice are software engineers with Attention Deficit Hyperactivity Disorder (ADHD), despite evidence of their presence in the developer population [4]–[6]. Software engineers with ADHD commonly experience challenges related to task initiation, sustained attention, organization, context switching, and task completion, which can affect their ability to maintain progress throughout software development activities [2], [3], [5], [7]–[9]. Although developers often adopt personal strategies such as timeboxing, reminders, body doubling, and structured task management, these approaches may be insufficient without support from tools and work environments. Formal accommodations also remain underused due to stigma, lack of awareness, and unclear processes [2], [3], [5], [7], [10].

At the same time, recent work has explored how digital tools and Large Language Models (LLMs) can support individuals with ADHD, primarily in clinical and everyday settings [11]–[15]. Research on programming support has similarly shown that LLMs can assist with coding and problem solving, while raising concerns regarding overreliance and reduced engagement when support focuses on direct answers [16]–[20]. However, existing ADHD-oriented tools do not address software development tasks, and existing LLM-based programming assistants do not account for the needs of neurodivergent developers. To address this gap, we investigate how LLM-based support can assist software engineers with ADHD in maintaining direction and progressing through development tasks. During our investigation of this problem space, we identified a previously published open-source system designed to support software engineers with ADHD [21]. Rather than developing a new system from scratch, we used the publicly available implementation as a foundation and extended it with support for planning, coding, debugging, and review, together with structured workflow guidance and persistent context management. We named the resulting system ***Tether 2.0*** to reflect its expanded functionality and workflow support. Our goal is to answer the following research question: **RQ. *How can an LLM-powered assistant be designed to support people with ADHD across planning, coding, debugging, and review in ways that support direction, sustained progress, and learning?***

To answer our RQ, we followed an engineering research approach centered on the design and evaluation of ***Tether 2.0***, a workflow-oriented system that supports planning, implementation, debugging, and review in software development. ***Tether 2.0*** incorporates structured interaction modes, persistent memory, activity-aware support, and workflow-specific assistance intended to help users maintain continuity across development tasks. We evaluated the system through a survey of domain experts and a semi-structured interview with a software engineer with ADHD following hands-on use. Data were analyzed using descriptive statistics and thematic analysis. Findings suggest that workflow-aware assistance supports task initiation, decomposition, and re-engagement while reducing cognitive overhead.

This paper makes three contributions. First, we present ***Tether 2.0***, an extension of an existing open-source ADHD support tool that introduces workflow-oriented assistance across planning, implementation, debugging, and review. Second, we provide a multi-source evaluation combining expert feedback and real-world usage, offering insights into how structured and context-aware support can assist software engineers with ADHD during development activities. Third, we contribute empirical evidence regarding the benefits and limitations of LLM-based workflow scaffolding, including its influence on cognitive load, distraction risk, task progression, and perceived usefulness in software engineering contexts.

## II. BACKGROUND

This section introduces the key concepts of the study, focusing on ADHD characteristics and technological support approaches, particularly in software engineering contexts.

### *A. ADHD in the Workplace Implications*

ADHD is a neurodevelopmental condition characterized by persistent patterns of inattention, hyperactivity, and impulsivity, but its effects extend beyond these core symptoms into everyday functioning [9], [22]. With an estimated 2.5% worldwide adult prevalence, ADHD is associated with difficulties in executive functioning, including planning, organization, time management, task initiation, working memory, and self-regulation, alongside impairments in academic, occupational, and social functioning [9], [23]–[25]. Recent reviews indicate that ADHD is linked to broader risks affecting mental health, physical health, and daily life outcomes, with implications for quality of life across the lifespan [24], [25]. These characteristics are directly relevant to complex work settings. In software engineering, developers with ADHD often report difficulties with sustained attention, task initiation and completion, planning, time management, and self-regulation [5], [7], [8]. These challenges are shaped not only by individual cognitive differences but also by the socio-technical environment, including frequent interruptions, context switching, synchronous communication, and the need to manage long and multi-stage tasks [3], [5], [8]. Although developers adopt coping strategies and may benefit from accommodations, existing support is often informal and not well integrated into development workflows [3], [8]. As a result, there is a gap between the characterization of ADHD-related challenges in software engineering and the design of tools that address them in practice.

### *B. Digital Tools for ADHD Support*

Existing digital tools for ADHD span educational, behavioral, and self-management domains. Reviews show that these tools commonly target academic learning, cognitive training, routine support, emotional regulation, and distraction management [14], [26]. Many applications focus on structuring activities, providing feedback, and supporting behavioral regulation through prompts, biofeedback, or sensory input [27]–[29]. Others target executive functioning more directly by restricting distractions, training working memory, or detecting disengagement and prompting re-engagement [28], [30]. Overall, these systems function as external scaffolds that support attention and task management, although they are primarily designed for educational or therapeutic contexts rather than professional work [14], [15]. In practice, these tools take several forms. Educational applications such as Say-it and Learn and Supangan use interactive feedback and gamification to support learning and engagement [27]–[29]. Tools such as TangiPlan, ChillFish, and BlurtLine support routines and self-regulation through structured guidance and feedback [28]. Cognitive training approaches include systems such as ADHD Trainer, which targets attention and memory [28], [30]. A subset of tools focuses specifically on attention regulation and distraction management, including Stayfocusd, Leechblock, SimplyNoise, N-back, Living Smart, Snappy, and CASTT, which use techniques such as distraction blocking, ambient feedback, and disengagement detection to support focus [28], [31]. Despite the range of approaches, these tools are rarely designed for complex professional domains such as software engineering. This limits their applicability to development work, where tasks are often long, ambiguous, and cognitively demanding. As a result, while existing tools demonstrate effective strategies for externalizing structure and supporting attention, they do not directly address the specific demands of software engineering workflows.

### *C. Tether: LLM-Based Support for Software Engineers with ADHD*

Tether is a desktop application designed to support software engineers with ADHD by providing context-aware assistance within development workflows [21]. The tool combines local activity monitoring, a conversational chatbot, retrieval-augmented generation, and gamification to address challenges related to attention, task initiation, and self-regulation. It uses signals such as active window usage and idle time to detect disengagement and deliver timely prompts, either through notifications or interactive dialogue. Its architecture is modular and primarily local, supporting privacy while enabling adaptive, personalized responses grounded in ADHD-related knowledge and prior interactions. Despite these contributions, the original version of Tether remains limited in scope and evaluation [21]. The system focuses primarily on attention support and task re-engagement, with less emphasis on guiding users through structured development activities such as planning, coding, debugging, and review. Its validation is preliminary and does not include studies with target users, limiting conclusions about its effectiveness in practice [21]. At the same time, its modular design and use of LLM-based components make it suitable for extension, providing a foundation for developing more structured and workflow-oriented support aligned with the demands of software engineering.

## III. METHOD

This study follows an Engineering Research approach focused on the design and evaluation of a software artifact [32]. During our investigation of ADHD support in software engineering, we identified a previously published open-source system designed to support software engineers with ADHD [21]. Rather than developing a new system from scratch, we used its publicly available implementation as a foundation. We retained key design foundations, including ADHD-informed principles that externalize executive-function demands [23], a local-first architecture, activity-aware prompting, conversational interaction, and retrieval-augmented generation [21]. Building on this foundation, we developed ***Tether 2.0***, a workflow-oriented extension that introduces explicit support for planning, implementation, debugging, and review. These additions address limitations identified in the original work, particularly the absence of structured guidance across multi-step development activities [21]. We also expanded the evaluation strategy beyond the preliminary validation reported for the original system [21]. The evaluation combined expert surveys, hands-on use by a software engineer with ADHD, and a follow-up semi-structured interview. This multi-source approach was designed to assess the system's perceived usefulness, relevance, and limitations while providing feedback from both domain experts and intended users. The following subsections describe the system design and evaluation procedures.

### A. Tool Development

*1) System Overview and Architecture:* ***Tether 2.0*** is implemented as a local-first desktop application designed to provide workflow-aware support within everyday programming activities. The system combines an Electron and React frontend with a Flask backend, a local SQLite database for persistent conversation storage, and a FAISS vector store for retrieval-augmented generation. Figure 1 presents the overall architecture, including the frontend, backend, persistence layer, retrieval component, and external model calls. This architecture supports conversational interaction while preserving continuity across sessions, an important consideration given prior research showing that conversational assistants can reduce interaction overhead and provide lightweight support within software engineering contexts [33], [34]. For language generation and retrieval embeddings, the system uses Gemini 2.5 Flash and Google's `embedding-001` model. All application logic, activity processing, session persistence, and document retrieval are executed locally, with only model calls sent externally. Retrieval-augmented generation was adopted to improve contextual grounding by allowing the system to draw on indexed ADHD-related reference material during response generation, consistent with prior work highlighting retrieval as a mechanism for improving relevance and response quality in LLM-based systems [35], [36]. Together, these components enable the system to combine persistent memory, retrieved support knowledge, and real-time developer context in a form aligned with software engineering workflows and ADHD-oriented scaffolding needs [16], [37], [38].

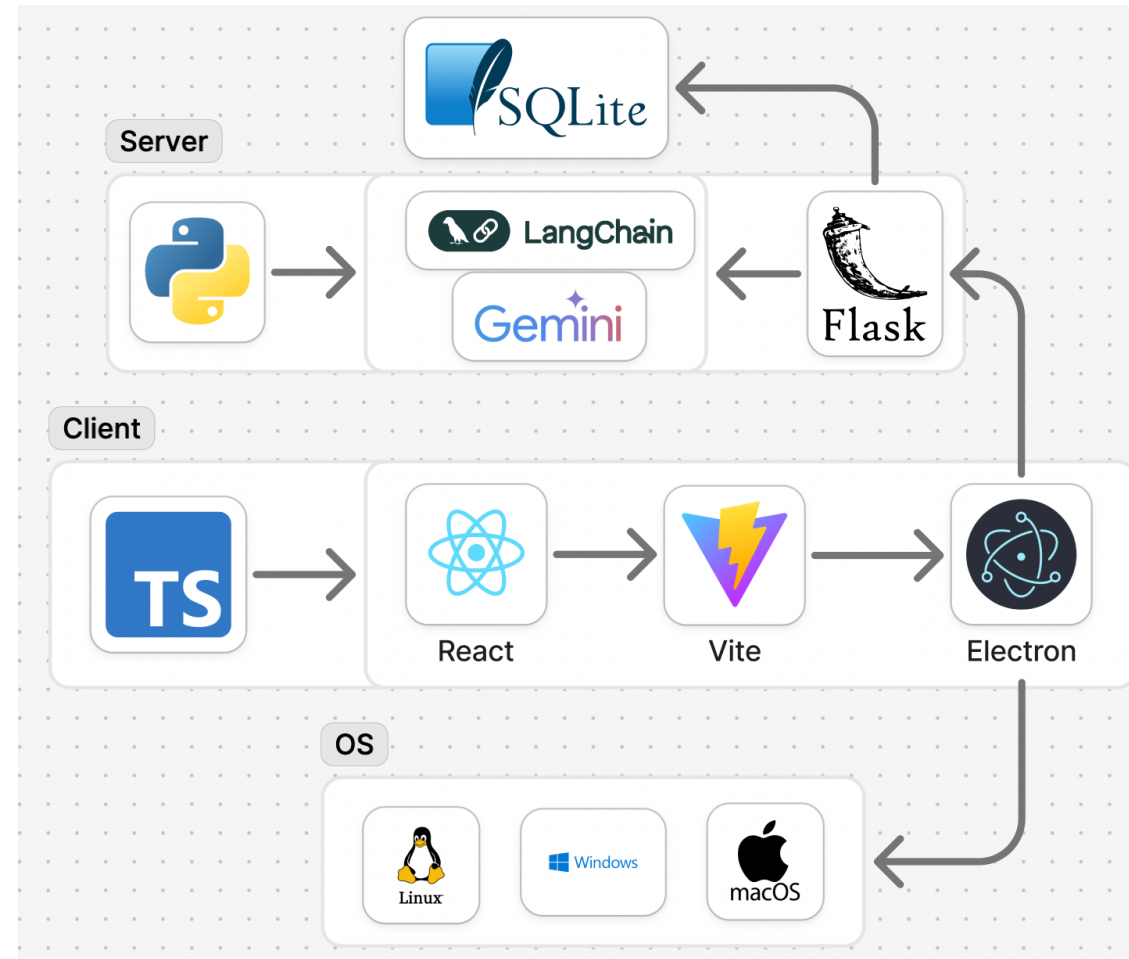

Fig. 1: ***Tether 2.0*** architecture including the frontend, backend, storage, retrieval, and LLM components.

*2) Activity Monitoring and Context Summarization:* ***Tether 2.0*** continuously captures lightweight behavioral signals from the local machine, including active window changes, application names, window titles, and idle-state transitions. Rather than passing raw logs directly to the language model, the backend aggregates them into structured summaries describing recent activity patterns, such as frequently used applications, prolonged inactivity, task switching, and uninterrupted focus periods. These summaries are injected into the system prompt, allowing responses to reflect recent work patterns rather than relying only on the current user message. This design draws on ADHD research emphasizing the value of externalizing executive-function demands into the environment [23] and software engineering studies identifying interruptions, context switching, and task-management difficulties as recurring challenges for developers with ADHD [3], [5], [8]. By converting activity traces into concise contextual signals, the system reduces the need for users to reconstruct task state manually and enables more timely workflow-aware support.

*3) Structured Workflow Design:* ***Tether 2.0*** introduces a structured workflow organized around four recurring stages of software development: planning, implementation, debugging, and review. Rather than relying on a single general-purpose conversational behavior, interactions are routed through four specialized modes, Planner, Builder, Detective, and Reviewer, each governed by a distinct system prompt while sharing the same session history. This design was motivated by ADHD-related needs for structure, decomposition, and stepwise guidance [23], [37], as well as evidence that developers with ADHD often encounter difficulties with task initiation, context switching, debugging frustration, and carrying work through to completion [3], [5], [8]. Planner mode elicits clarifying information about project goals, scope, and technical decisions before generating a time-bounded checklist of implementation tasks (Figure 2 and Figure 3).

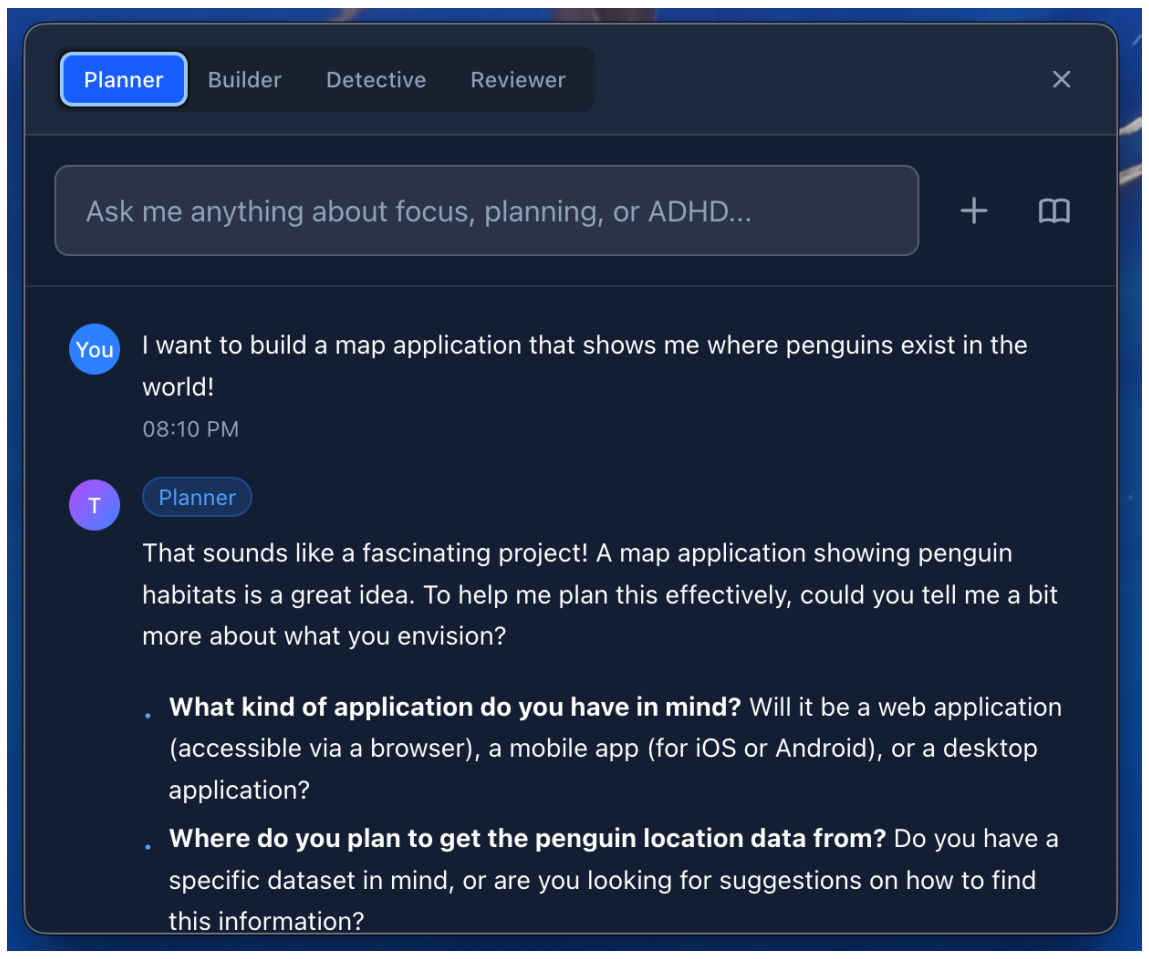

Fig. 2: Planner mode initiating the structured workflow through clarifying dialogue.

Builder mode supports implementation by explaining concepts, generating commented examples, and encouraging incremental progress (Figure 4). Detective mode assists debugging through guided questioning or direct-fix strategies, allowing the assistant to act either as a teaching companion or a corrective support mechanism (Figure 5). Reviewer mode focuses on completion by helping users verify functionality, create lightweight tests and documentation, identify refactoring opportunities, and reflect on lessons learned (Figure 6). Together, these modes position ***Tether 2.0*** as a workflow-oriented scaffold designed to support both ADHD-related needs and broader software engineering activities requiring structured guidance [37], [38].

*4) Retrieval, Memory, and Checklist Persistence:* ***Tether 2.0*** combines retrieval-augmented generation with persistent session memory and lightweight task persistence. The system maintains a local FAISS vector store containing ADHD-related reference materials, including literature on emotional regulation, self-regulation, and workplace experiences. Retrieved content can be incorporated dynamically during response generation to improve contextual relevance and grounding [35], [36]. All user and assistant messages are stored in a local SQLite-backed conversation history and tagged according to interaction mode. Although the modes are behaviorally distinct, they share a common session history, allowing planning discussions to inform later implementation, debugging, and review activities. This shared-memory design is particularly relevant in ADHD-related support settings, where repeatedly reconstructing task state can increase cognitive overhead [23], [38]. In addition, Planner-generated task lists are persisted as structured checklists that can be revisited and updated over time. Together, retrieval, shared memory, and checklist persistence allow the system to function as a longitudinal workflow scaffold that preserves context across multiple stages of software development.

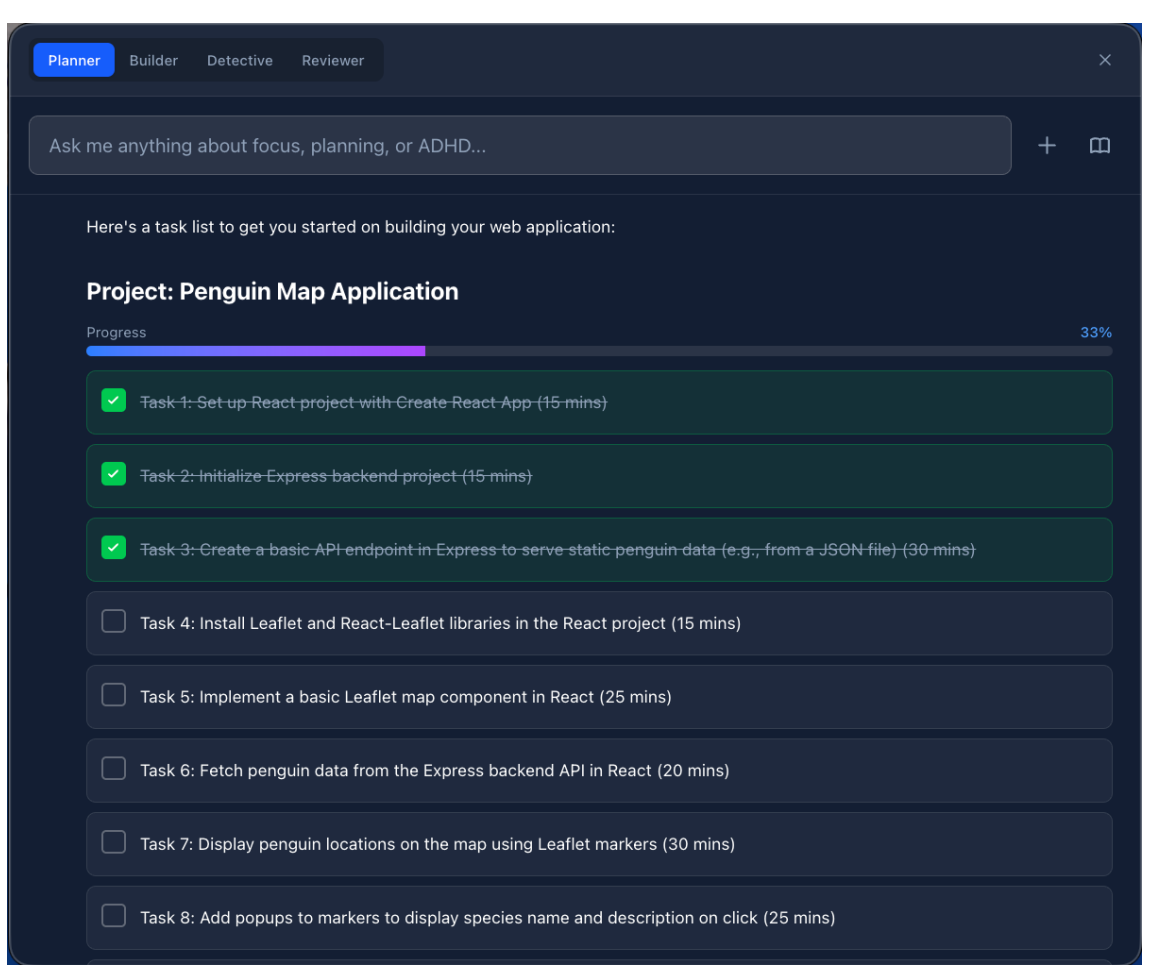

Fig. 3: Structured checklist example generated by Planner mode to support task decomposition and progress tracking.

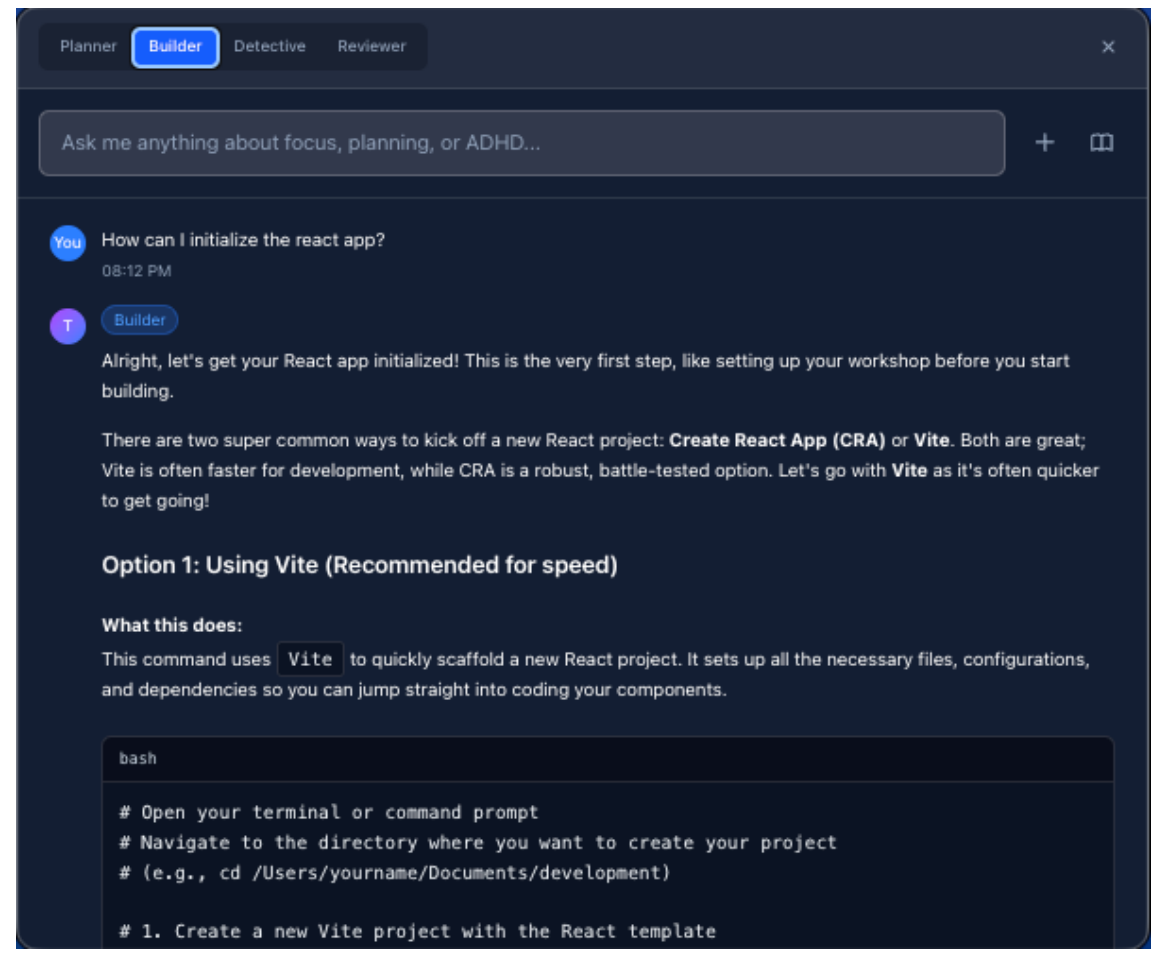

Fig. 4: Builder mode supporting set up efforts for the project while explaining executed commands.

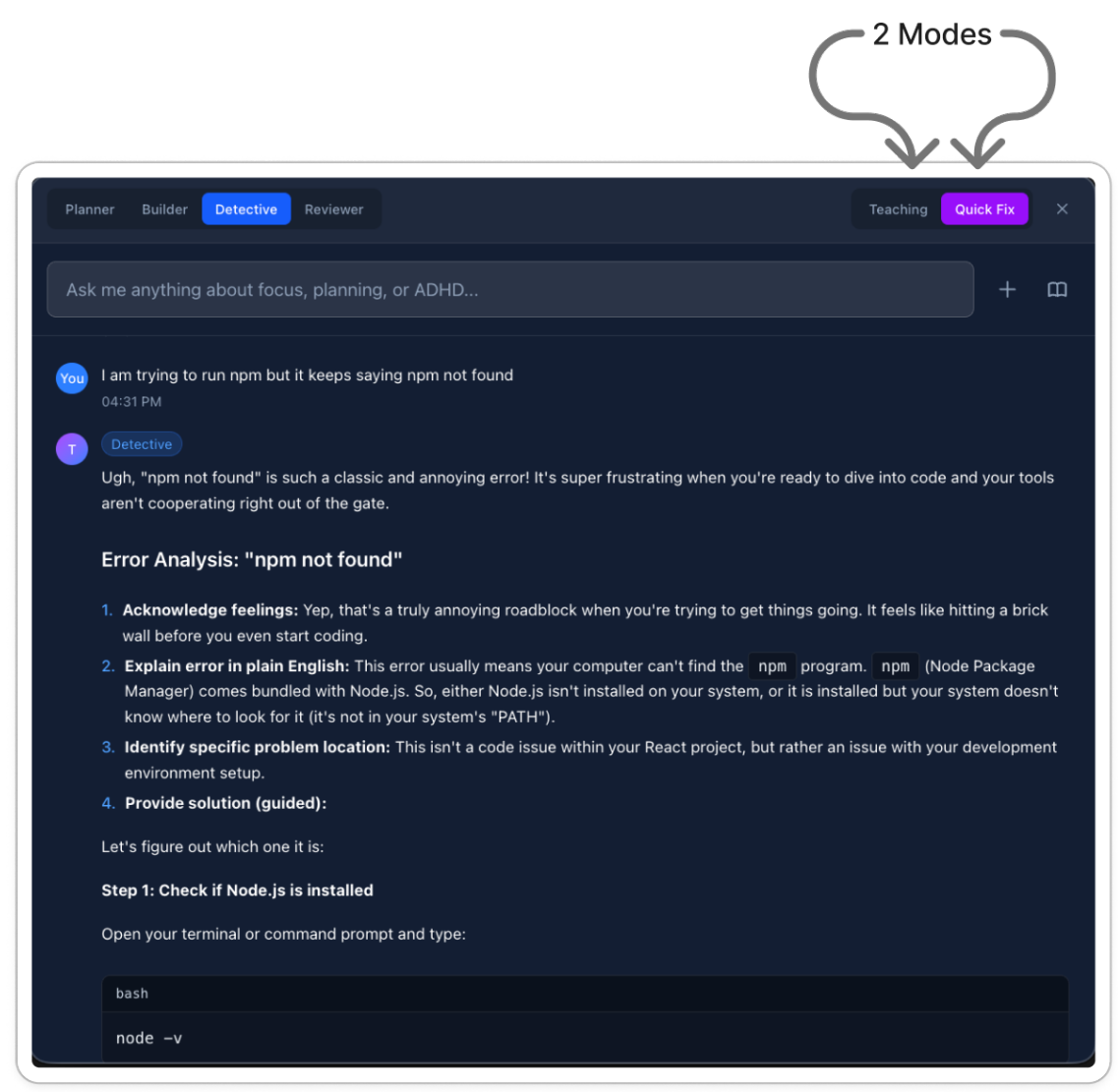

Fig. 5: Detective mode supporting debugging through guided questioning and direct-fix strategies.

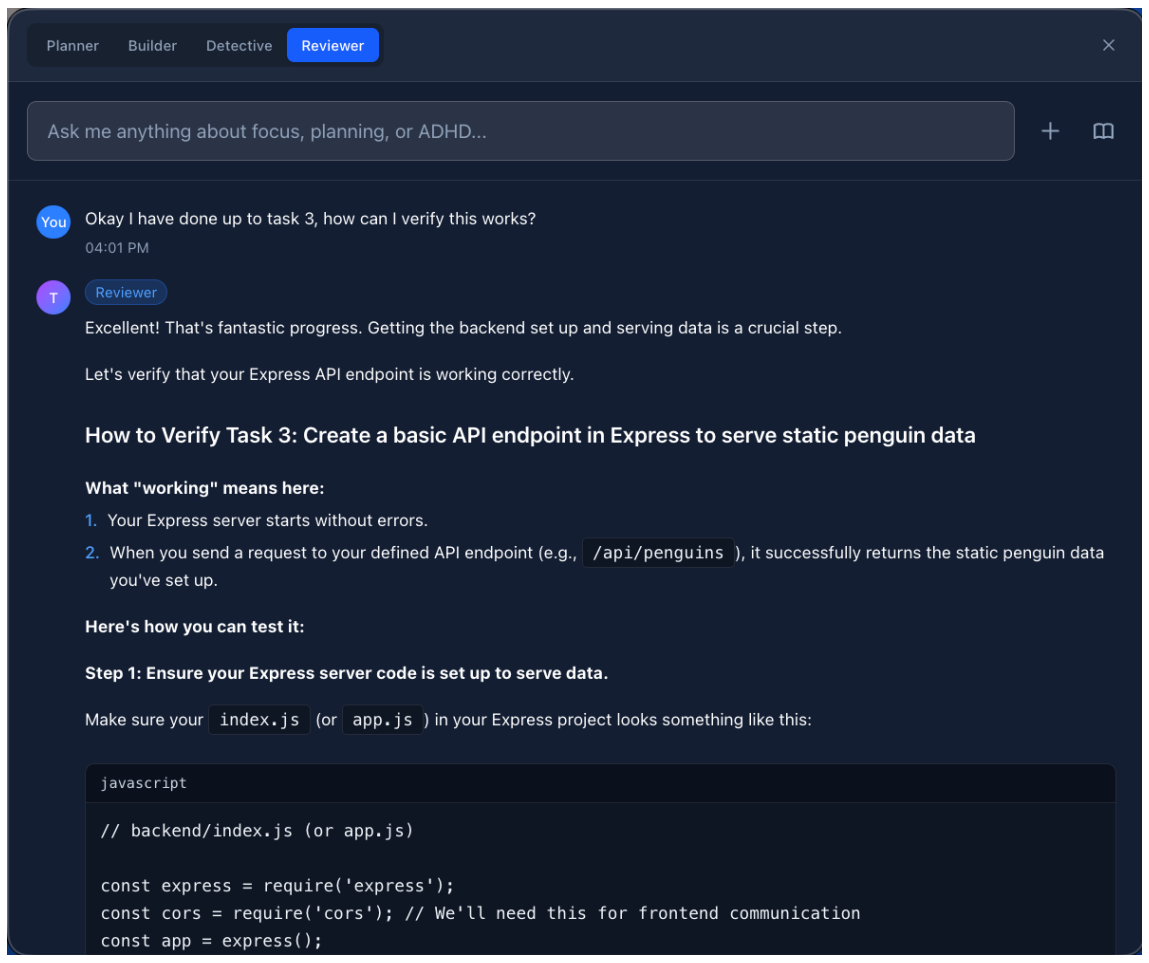

Fig. 6: Reviewer mode showing assistance during review of a previously completed task.

### B. Tool Evaluation

***Tether 2.0*** was evaluated through a multi-source process focused on utility, workflow fit, and domain relevance. Consistent with engineering-oriented evaluation practices in software engineering research [32], we combined expert surveys with hands-on use and interview-based feedback from a software engineer with ADHD. This strategy enabled us to assess the artifact from both expert and end-user perspectives while identifying opportunities for refinement.

*1) Survey-Based Evaluation:* We conducted a survey-based evaluation to assess the perceived usefulness, relevance, and design adequacy of ***Tether 2.0***. Although software engineers with ADHD were the intended target population, recruiting this group proved challenging. Therefore, we expanded recruitment to include participants with expertise in ADHD and software engineering who could provide informed assessments of the tool's design and applicability. Participants were recruited using purposive and snowball sampling [39]. Three groups were targeted: (1) clinicians, psychologists, and ADHD-focused researchers affiliated with the authors' institution; (2) software engineering researchers studying ADHD, identified through Google Scholar searches using the terms *Software Engineering* and *ADHD*; and (3) researchers with broader expertise in ADHD. This strategy was intended to capture perspectives related to clinical plausibility, ADHD relevance, and software engineering practice [3], [5], [8], [37]. All participants received a recorded demonstration video and the survey instrument. The video presented the system architecture, workflow modes, and design rationale. Following recommendations for empirical software engineering research [32], [40], we developed a study-specific instrument composed of open-ended questions and five-point Likert-scale items. A pilot with two software engineering students resulted

TABLE I: Expert Validation Instrument

**OPEN-ENDED QUESTIONS**

1) From your perspective, how useful could this tool be for supporting task initiation in software engineers with ADHD during real development work? Why?
2) Which software-engineering-specific ADHD challenges (e.g., starting new work, handling large codebases, debugging frustration, finishing tasks) do you think this tool addresses best?
3) In which software engineering contexts (e.g., junior developers, senior developers, high-interrupt teams, remote work, startup environments) could this tool be less effective or potentially problematic?
4) If you could change or add one feature to improve usefulness for software engineers with ADHD in real workflows, what would it be?
5) How do you think a structured workflow assistant like this could change how software engineers with ADHD approach development tasks?

**SCALE-BASED ITEMS**

6) How well could this tool support starting a development task when a software engineer with ADHD is stuck or avoiding initiation? (5-point scale: 1 = Not at all, 5 = Extremely well)
7) How well could this tool support breaking down large coding tasks into actionable steps? (5-point scale: 1 = Not at all, 5 = Extremely well)
8) How well could this tool reduce cognitive overwhelm when beginning complex engineering work (e.g., a new codebase, a large work item, or unfamiliar technologies)? (5-point scale: 1 = Not at all, 5 = Extremely well)
9) How well could this tool support sustained focus during long coding or debugging sessions? (5-point scale: 1 = Not at all, 5 = Extremely well)
10) How well could this tool support re-engagement after distraction or context switching? (5-point scale: 1 = Not at all, 5 = Extremely well)
11) How well could this tool support finishing tasks (e.g., tests, documentation, refactoring, and definition of done)? (5-point scale: 1 = Not at all, 5 = Extremely well)
12) How likely is it that software engineers with ADHD would adopt a tool like this in daily development work? (5-point scale: 1 = Not at all likely, 5 = Extremely likely)
13) How likely is this tool to become distracting or cognitively noisy over time? (5-point scale: 1 = Not at all likely, 5 = Extremely likely) in minor wording refinements. Open-ended questions explored usefulness, addressed challenges, limitations, and potential improvements, while the scale items assessed support for task initiation, task decomposition, reduction of cognitive overwhelm, sustained focus, re-engagement after distraction, task completion, likelihood of adoption, and potential distraction risk. The complete instrument is presented in Table I.

*2) Interview-Based Feedback:* To complement expert feedback, we sought direct input from the intended user population. We recruited one software engineer with ADHD who used ***Tether 2.0*** in their own environment for approximately two weeks before participating in a semi-structured interview. The participant received instructions for installation and use, allowing the discussion to be grounded in sustained interaction with the system rather than a conceptual walkthrough. The interview lasted approximately 25 minutes and explored experiences with the tool, perceived benefits, workflow fit, and opportunities for improvement. The discussion reused the core survey questions while incorporating follow-up prompts to obtain richer feedback regarding usability and real-world applicability. This approach maintained consistency with the broader evaluation while capturing lived experience from an intended user.

*3) Data Analysis:* Survey and interview data were analyzed using complementary qualitative and quantitative approaches. Open-ended survey responses and interview transcripts were analyzed through thematic analysis [41]. The analysis focused on perceptions related to ADHD-specific software engineering challenges, including task initiation, decomposition, sustained focus, re-engagement after interruptions, and task completion. Responses were coded, grouped into categories, and synthesized into higher-level themes describing participants' views on the role of structured LLM-based assistance in software development. Likert-scale responses were analyzed using descriptive statistics, including means, to summarize participant assessments across the evaluated dimensions. These quantitative results complement the qualitative findings by providing an overview of perceptions regarding the system's core design goals.

*4) Ethical Considerations:* We obtained approval from the institutional research ethics board prior to conducting this study. Participants were informed about the study objectives, data usage, and voluntary nature of participation, and consent was obtained before data collection. All procedures followed recommended ethical practices for empirical software engineering research, including participant anonymity, secure data handling, and responsible reporting [32].

## IV. RESULTS

This section presents the results of this study, describing the ***Tether 2.0***, its behavior in practice, and the evaluation conducted with knowledge users and domain experts. To support tool use, we made a video available on https://figshare.com/s/60c908e4475cf16f5b91.

### A. Tether 2.0 User Workflow in Practice

Tether 2.0 supports software development through four modules that correspond to common development activities: Planner, Builder, Detective, and Reviewer. Although developers can enter any module at any time, a typical workflow begins with planning, progresses through implementation and debugging, and concludes with review activities. Throughout this process, conversations, activity summaries, and task checklists remain available, allowing work to continue across sessions without requiring users to reconstruct context.

A typical workflow is as follows:

- **Planning.** The developer begins by describing a task, feature, or problem. Planner asks clarifying questions regarding requirements, constraints, and implementation choices, then generates a structured checklist that decomposes the work into smaller actionable tasks.
- **Implementation.** The developer moves to Builder and works through the generated checklist. Builder provides explanations, code examples, setup assistance, and implementation guidance while preserving awareness of the overall task plan.
- **Debugging.** If problems arise during implementation, the developer can switch to Detective. This module either guides the developer through a structured debugging process or provides direct suggestions regarding likely causes and potential fixes. Once the issue is resolved, development can continue in Builder.
- **Review and completion.** After implementation is finished, the developer transitions to Reviewer. This module assists with verification, lightweight testing, documentation, code review activities, and identification of remaining tasks before completion.
- **Resuming interrupted work.** At any point, developers can leave and later return to a task. Previous conversations, activity summaries, and checklist artifacts are preserved, allowing the system to provide recaps and suggest appropriate next steps.

This workflow allows developers to move between planning, implementation, debugging, and review while maintaining a shared context. Rather than acting as an isolated coding assistant, Tether 2.0 functions as a workflow companion that supports task decomposition, progression, and re-engagement throughout the software development process.

### B. Tether 2.0 Evaluation

Our evaluation involved sixteen participants with diverse expertise spanning software engineering, human–computer interaction, psychology, and industry practice. The sample was structured to capture complementary perspectives on clinical relevance, software engineering applicability, and practical use. Participants were grouped into four primary buckets. The largest subgroup consisted of software engineering researchers (6/16), including doctoral students, PhD candidates, and associate professors working on developer experience, human factors, accessibility, and ADHD-related topics in software engineering. The healthcare and psychology subgroup (4/16) included clinical psychology and neuropsychology faculty, as well as researchers and practitioners engaged in ADHD-focused research. The computing and HCI subgroup (3/16) comprised faculty members and researchers with expertise in interaction design, user experience, and system evaluation. Finally, industry and early-career computing participants (3/16) included an industry mentor with experience in AI systems and product development, as well as early-career practitioners and researchers with experience in software development and tooling. In terms of seniority, the sample included professors (5/16), doctoral-level researchers or PhD candidates (6/16), and early- to mid-career participants (5/16). This distribution reflects representation across different levels of experience, from emerging researchers to established experts. Participants were geographically distributed across Canada (4/16), Mexico (3/16), Brazil (4/16), Portugal (1/16), United Kingdom (1/16), India (1/16), Sri Lanka (1/16), and Iceland (1/16). With respect to gender, the sample included women (9/16) and men (7/16). Additionally, participants with disclosed ADHD (2/16) and participants without disclosed ADHD (14/16) were represented. Overall, the sample combines expertise in software engineering, ADHD research, and industry practice, enabling assessment of the tool from multiple relevant perspectives, including clinical plausibility, workflow alignment, and anticipated usability.

*1) Scale-Based Evaluation:* The scale-based items assessed participants' perceptions of how well the tool could support key stages of software engineering work commonly associated with ADHD-related challenges, as well as its potential adoption and the associated cognitive risks. Responses were collected using a five-point Likert scale (1 = Not at all, 5 = Extremely well). Ratings indicate a gradient of perceived support across variables, from higher to lower. The highest ratings were observed for *task decomposition* (mean = 4.00), followed by *task initiation* (mean = 3.81) and *cognitive overwhelm reduction* (mean = 3.75). These results suggest that participants perceive the tool as particularly effective in structuring work and supporting entry into complex development tasks. Moderate ratings were observed for *re-engagement after distraction* (mean = 3.63), indicating that the tool may also support recovery from context switching, although to a lesser extent than initial task structuring. Lower ratings were reported for *sustained focus* (mean = 3.31) and *task completion* (mean = 3.31), suggesting less perceived support for maintaining attention over extended periods and for completing tasks. With respect to practical use, the *perceived distraction risk* (mean = 3.19) slightly exceeded the *likelihood of adoption* (mean = 3.13), both of which were rated in the moderate range. This pattern indicates that participants identified potential cognitive costs associated with the tool, which may influence their willingness to adopt it in everyday workflows. Overall, the distribution of ratings suggests that the tool is perceived as more effective at supporting early-stage and structuring activities, while its contribution to sustained engagement and completion is viewed as comparatively weaker. The moderate evaluations of adoption and cognitive risk further suggest that integrating it into real development contexts may depend on how effectively these trade-offs are addressed.

*2) Qualitative Evaluation:* Using thematic analysis, we organized participants' narratives into four themes: *Perceived Benefits*, *Perceived Limitations and Risks*, *Perceived Mechanisms of Support*, and *Contextual Fit and Moderating Factors*. These themes capture how participants interpreted the tool's role in supporting software engineering work involving ADHD-related challenges.

**Perceived Benefits.** Aligned with the scale-based items, participants described the tool as particularly beneficial for supporting *task initiation* and *task decomposition*. Many emphasized its role in reducing the difficulty of starting work by transforming large or abstract tasks into manageable steps. One participant noted that *"the main difficulty is often not knowing where to begin, and this tool seems to turn large, vague development tasks into smaller, actionable starting points" (P07)*. Similarly, another highlighted that *"a quick run-through of the content may help users by outlining low hanging fruit, and easy to complete tasks that will help them engage with the work without initial overwhelm" (P06)*. The tool was also perceived as reducing cognitive burden during early stages of development. Participants described benefits related to structuring and organization, with one stating that it *"can help developers overcome the initial barrier and start their tasks more easily" (P11)* and another noting that it supports *"planning and organization of the steps required to complete a given activity" (P15)*. In addition, participants identified value in supporting re-engagement after interruptions, particularly through reminders and workflow tracking. For instance, one participant observed that *"the notifications system also seems to be a good addition which addresses the issue of distraction" (P07)*.

**Perceived Limitations and Risks.** Despite these benefits, participants raised several concerns regarding potential limitations and risks. A recurring issue was the possibility that the tool could increase cognitive load or become distracting over time. One participant cautioned that *"it might provide an additional distraction for users (i.e., another app to interact with/detract from work)" (P03)*, while another emphasized that *"such tools can actually cause cognitive overload" (P05)*. Participants also highlighted risks related to over-reliance on automated support, particularly for less experienced developers. One participant noted that *"junior devs may be more likely to implement the provided suggestions and less likely to ask a senior dev for help" (P06)*, while another warned that *"they may develop a dependency on the tool and may not be motivated enough to learn how to solve the problem" (P09)*. Concerns were also raised about the role of generative AI, with one participant stating that *"we are walking on very thin ice with these AI assistants" (P02)*, pointing to broader risks associated with cognitive offloading.

**Perceived Mechanisms of Support.** Participants' accounts also provide insight into the mechanisms through which the tool is perceived to support software engineering work. A central mechanism is the structuring of tasks through decomposition and guided workflows. One participant described this as *"going from 'I need to do big thing' to 'here are the little things, and this is the next one' " (P08)*, highlighting the role of task chunking. Another emphasized the importance of structured interaction, noting that *"the structured guidance, notifications, and clear workflow can reduce distractions and make it easier to keep attention on the current development task" (P07)*. Additional mechanisms include reminders and context recovery. Participants pointed to features such as notifications and idle detection as ways to support re-engagement, with one explaining that *"tracking when the user becomes inactive and which application or window is active seems very relevant for ADHD-related focus management" (P14)*. These mechanisms suggest that the tool operates not only by providing information, but by scaffolding interaction with tasks over time.

**Contextual Fit and Moderating Factors.** Participants further emphasized that the effectiveness of the tool depends on contextual and individual factors. Differences in experience level were frequently mentioned, particularly with respect to junior developers. One participant noted that *"for junior devs,* *this tool may be problematic by replacing critical thinking and learning with easy to access solutions" (P06)*. Similarly, another observed that *"more experienced developers may already have their own strategies to organize tasks and manage attention" (P11)*. The work environment was also identified as an important moderating factor. Participants suggested that the tool may be more effective in settings with fewer interruptions, with one stating that *"remote work across junior and senior devs is the best place to use this tool" (P05)*, while another cautioned that *"in environments with excessive interruptions, the tool may unintentionally intensify difficulties related to maintaining focus" (P15)*. Finally, individual differences in ADHD experience and preferences were highlighted. One participant emphasized that *"ADHD is very individual - what works for one individual might not work for another" (P16)*, underscoring the importance of customization and adaptability. These observations indicate that the perceived usefulness of the tool is contingent on both situational and personal factors, rather than being uniform across users and contexts.

*3) Target User Evaluation: Two-Week Use and Interview:* To complement the survey-based and scale-based evaluations, we conducted a two-week deployment with a software engineer with ADHD, followed by a semi-structured interview. This evaluation provides a perspective grounded in sustained interaction and lived experience, allowing us to compare how the perceived benefits, limitations, and mechanisms identified by knowledgeable participants manifest in practice. The participant reported a predominantly positive experience with the tool, emphasizing its usefulness in supporting *task initiation*, *task decomposition*, and awareness of work patterns. He noted that *"being able to split work into smaller pieces and check off progress would help me start tasks more easily" (P-U1)*, and highlighted that the system helped him monitor his own behavior, stating that *"it helped me realize when I was being productive or distracted" (P-U1)*. Notifications were also perceived as effective for re-engagement, as *"the positive notifications actually pushed me back to work when I lost focus" (P-U1)*. While some minor issues were mentioned, such as occasional notification failures, these did not substantially affect the overall perception of usefulness.

The participant identified the planner as the most valuable component, explaining that *"I struggle a lot with breaking tasks into smaller chunks and managing multiple responsibilities, so a planner like this could help me a lot" (P-U1)*. He also emphasized that the tool supported both starting and finishing tasks, which contrasts with the more cautious assessment from knowledgeable participants, who rated *task completion* and *sustained focus* more moderately. At the same time, he suggested improvements related to activity monitoring, proposing that *"if I spend too long on something unrelated, it could prompt me to return to work" (P-U1)*, indicating a desire for stronger support in managing attention over time.

Some aspects of the experience extend beyond what was explicitly emphasized in the survey and scale-based evaluations. In particular, the participant highlighted the tool's role in increasing awareness of productivity patterns and in supporting temporal planning. He explained that *"it would help me break tasks into chunks and plan my day better, especially around meetings. I often wait around for meetings and lose productivity" (P-U1)*, suggesting that the tool may contribute to reflective awareness and coordination of work across time. Contextual factors were also relevant, as he noted that *"being in a non-work environment makes it harder to start working" (P-U1)*, indicating that the perceived usefulness of the tool may be amplified in remote or distraction-prone environments.

## V. DISCUSSION

This section discusses the findings by situating the results within existing literature, outlining implications for research and practice, and addressing threats to validity.

### *A. Comparing Results with the Literature*

Our results align with prior ADHD and digital support research, which identifies external scaffolds, task structuring, re-engagement prompts, and self-regulation support as central mechanisms for assisting individuals with ADHD [14], [26]–[28], [30]. Consistent with these findings, Tether 2.0 externalizes task decomposition through checklists, preserves interaction history, and incorporates activity summaries to support re-engagement. These mechanisms reflect established approaches to reducing cognitive load and supporting executive functioning [23]. Our evaluation provides support for these mechanisms, particularly regarding *task initiation*, *task decomposition*, and *cognitive overwhelm reduction*. However, both the survey and interview findings suggest more moderate support for *sustained focus* and *task completion*, indicating that external scaffolding may be more effective for entering and organizing work than for maintaining long-term engagement.

The findings also align with software engineering studies describing ADHD-related challenges such as task initiation, sustained attention, context switching, and task completion in interruption-prone environments [3], [5], [7], [8]. Unlike many ADHD-focused tools, which are not designed for complex professional activities or multi-stage tasks [14], [15], Tether 2.0 integrates support directly into software development workflows. Extending both existing ADHD support approaches and the original Tether artifact [21], the system supports planning, implementation, debugging, and review within a shared workflow. Similar to recent research on LLM-based learning and programming support [17]–[20], our findings suggest that the greatest value comes from decomposition, reasoning, and guided progression rather than direct answer generation. At the same time, participants raised concerns consistent with the literature regarding over-reliance, reduced critical engagement, and inappropriate levels of assistance for less experienced developers.

The target user evaluation further refined these findings. While knowledgeable participants expressed moderate expectations regarding sustained focus, completion, and adoption, the software engineer with ADHD reported consistently positive experiences across these dimensions. This suggests that direct use may mitigate some anticipated limitations, particularly when support is embedded within real workflows. Nevertheless, both evaluation sources indicate that effectiveness depends on contextual factors such as work environment, developer experience, and integration with existing tools. In response to our research question, the findings suggest that LLM-based systems can support software engineers with ADHD primarily through structured workflow scaffolding and attention-aware interaction, with effectiveness varying across stages of work and conditions of use.

### B. Implications

Our work contributes to research on diversity in software engineering by advancing the translation of neurodiversity-related insights into the design of development tools. Rather than re-establishing known challenges, the findings suggest that integrating workflow-oriented support directly into development environments is a promising direction for operationalizing ADHD support in practice. Our results also indicate that effectiveness cannot be assumed as uniform across tasks or users, pointing to the need for models that account for variability in interaction patterns, experience levels, and work contexts. In addition, the presence of both supportive and potentially constraining effects highlights the importance of studying these systems as socio-technical interventions, where outcomes depend on how guidance is configured, interpreted, and incorporated into existing work practices. Future research may build on this by examining how such systems evolve with continued use, how users calibrate reliance over time, and how design choices influence both short-term behavior and longer-term skill development.

For industry practice, introducing support for attention and task management within development workflows requires careful consideration of how such support is delivered and integrated. Tools should prioritize alignment with existing environments and minimize additional interaction demands, as friction can limit practical uptake. One-size-fits-all solutions are unlikely to be effective, making adaptability to individual preferences and work conditions an important requirement. Contextual factors, such as work setting and experience level, also shape how such tools are used, suggesting that deployment should account for variation across teams and environments. These considerations point to the importance of designing systems that can be incrementally adopted, configured, and adjusted to support diverse work practices.

### C. Threats to Validity

The findings should be interpreted in light of several limitations related to the study design and the scope of evaluation. The evaluation remains preliminary and does not provide evidence of long-term effectiveness in real-world development settings. Much of the feedback was based on demonstrations and short-term interactions rather than sustained use under typical industry conditions such as interruptions, deadlines, and team coordination. The study combines multiple sources of evidence with varying levels of ecological validity, including surveys, demonstrations, and limited longitudinal interaction, resulting in uneven coverage across system features and potential novelty effects. In addition, the evaluation focuses on perceived utility and workflow fit rather than direct measurement of outcomes such as productivity, learning, or well-being, meaning that positive perceptions do not necessarily indicate long-term benefits. The study follows an engineering research approach, in which novel artifacts are evaluated through a combination of conceptual and empirical methods rather than through large-scale controlled studies [32]. The goal is to assess the artifact's plausibility, usefulness, and limitations rather than to establish causal effects. The participant pool was small and purposively selected, which limits generalization but aligns with qualitative research practices in software engineering and provides analytically grounded insights from participants selected for their relevance [42]. There are also system-related validity concerns, as LLM-based responses may be variable, incomplete, or mismatched to user needs, and features intended to support attention may introduce cognitive overhead for some users. Finally, applicability may vary across technical and organizational contexts, indicating that the results represent an initial step rather than a comprehensive validation.

## VI. Conclusions

In this study, we investigated how to design an LLM-powered assistant to support software engineers with ADHD across planning, coding, debugging, and review. We addressed the gap between documented ADHD-related challenges in software engineering and the lack of workflow-integrated tools by extending a prior system into Anonymous Tool 2.0, a structured, workflow-oriented assistant. Our results suggest that Anonymous Tool 2.0 supports development through interaction modes aligned with key stages: Planner supports requirement clarification and task decomposition, Builder assists with incremental coding, Detective supports debugging through guided reasoning or direct suggestions, and Reviewer assists with validation, testing, and documentation. Across these stages, the system preserves context and interaction history, enabling users to resume work and maintain progress without having to reconstruct the task state. Future work will focus on longer-term in situ deployment with larger and more diverse populations, incorporating longitudinal and outcome-based evaluation to assess real-world impact. Additional directions include deeper personalization, refinement of instructional and debugging support, integration of richer contextual signals, and broader validation across technical and organizational settings to better understand how such systems can be incorporated into everyday software engineering environments.

**Data availability Statement:** The replication materials are available as follows: Code: https://figshare.com/s/b57f49f92ff61cb81298. Survey and evaluation data: https://figshare.com/s/713a7d8be307c00d1f64?file=63055672. Tool demo: https://figshare.com/s/60c908e4475cf16f5b91

## REFERENCES

[1] K. Albusays, P. Bjorn, L. Dabbish, D. Ford, E. Murphy-Hill, A. Serebrenik, and M.-A. Storey, "The diversity crisis in software development," *IEEE Software*, vol. 38, no. 2, pp. 19–25, 2021.

[2] N. da Silva Menezes, T. Á. da Rocha, L. S. S. Camelo, and M. P. Mota, ""i felt pressured to give 100% all the time": How are neurodivergent professionals being included in software development teams?" in *Simpósio Brasileiro de Sistemas de Informação (SBSI)*. SBC, 2025, pp. 525–534.

[3] K. Newman, S. Snay, M. Endres, M. Parikh, and A. Begel, ""get me in the groove": A mixed methods study on supporting ADHD professional programmers," in *2025 IEEE/ACM 47th Intl Conference on Software Engineering (ICSE)*. IEEE Computer Society, 2025, pp. 778–778.

[4] Stack Overflow, "Stack overflow developer survey 2022," 2022, accessed: 2025-06-24. [Online]. Available: https://survey.stackoverflow.co/2022/

[5] K. Gama, G. Liebel, M. Goulão, A. Lacerda, and C. Lacerda, "A socio-technical grounded theory on the effect of cognitive dysfunctions in the performance of software developers with ADHD and autism," in *2025 IEEE/ACM 47th International Conference on Software Engineering: Software Engineering in Society (ICSE-SEIS)*. IEEE, 2025, pp. 1–12.

[6] P. Verma, M. V. Cruz, and G. Liebel, "Differences between neurodivergent and neurotypical software engineers: Analyzing the 2022 stack overflow survey," *arXiv preprint arXiv:2506.03840*, 2025.

[7] M. R. Morris, A. Begel, and B. Wiedermann, "Understanding the challenges faced by neurodiverse software engineering employees: Towards a more inclusive and productive technical workforce," in *Proceedings of the 17th International ACM SIGACCESS Conference on computers & accessibility*, 2015, pp. 173–184.

[8] G. Liebel, N. Langlois, and K. Gama, "Challenges, strengths, and strategies of software engineers with adhd: A case study," in *Proceedings of the 46th International Conference on Software Engineering: Software Engineering in Society*, 2024, pp. 57–68.

[9] S. Cortese, M. A. Bellgrove, I. Brikell, B. Franke, D. W. Goodman, C. A. Hartman, H. Larsson, F. R. Levin, E. G. Ostinelli, V. Parlatini *et al.*, "Attention-deficit/hyperactivity disorder (adhd) in adults: evidence base, uncertainties and controversies," *World psychiatry*, vol. 24, no. 3, pp. 347–371, 2025.

[10] R. d. S. Santos, C. Magalhaes, R. Santos, and J. Correia-Neto, "Exploring hybrid work realities: A case study with software professionals from underrepresented groups," in *Companion Proceedings of the 32nd ACM International Conference on the Foundations of Software Engineering*, 2024, pp. 27–37.

[11] S. Berrezueta-Guzman, M. Kandil, M.-L. Martín-Ruiz, I. Pau de la Cruz, and S. Krusche, "Future of ADHD care: Evaluating the efficacy of chatgpt in therapy enhancement," in *Healthcare*, vol. 12, no. 6. MDPI, 2024, p. 683.

[12] S. Berrezueta-Guzman, M. Kandil, and S. Wagner, "Integrating ai into ADHD therapy: Insights from chatgpt-4o and robotic assistants," *Human-Centric Intelligent Systems*, pp. 1–16, 2025.

[13] Y. Bannett, F. Gunturkun, M. Pillai, J. E. Herrmann, I. Luo, L. C. Huffman, and H. M. Feldman, "Applying large language models to assess quality of care: Monitoring ADHD medication side effects," *Pediatrics*, vol. 155, no. 1, p. e2024067223, 2025.

[14] E. Gabarron, K. Denecke, and G. Lopez-Campos, "Evaluating the evidence: a systematic review of reviews of the effectiveness and safety of digital interventions for adhd," *BMC psychiatry*, vol. 25, no. 1, p. 414, 2025.

[15] F. J. Schofield, S. Wilkie, E. E. Nielsen, A. Brigden, M. W. Jones, and H. K. Isotalus, "Digital health technologies for adults with adhd: a scoping review," *Frontiers in Digital Health*, vol. 8, p. 1746732, 2026.

[16] A. Fan, B. Gokkaya, M. Harman, M. Lyubarskiy, S. Sengupta, S. Yoo, and J. M. Zhang, "Large language models for software engineering: Survey and open problems," in *2023 IEEE/ACM International Conference on Software Engineering: Future of Software Engineering (ICSE-FoSE)*. IEEE, 2023, pp. 31–53.

[17] J. Prather, B. N. Reeves, P. Denny, B. A. Becker, J. Leinonen, A. Luxton-Reilly, G. Powell, J. Finnie-Ansley, and E. A. Santos, ""it's weird that it knows what i want": Usability and interactions with copilot for novice programmers," *ACM transactions on computer-human interaction*, vol. 31, no. 1, pp. 1–31, 2023.

[18] J. Prather, B. N. Reeves, J. Leinonen, S. MacNeil, A. S. Randrianasolo, B. A. Becker, B. Kimmel, J. Wright, and B. Briggs, "The widening gap: The benefits and harms of generative ai for novice programmers," in *Proceedings of the 2024 ACM Conference on International Computing Education Research-Volume 1*, 2024, pp. 469–486.

[19] M. Kazemitabaar, O. Huang, S. Suh, A. Z. Henley, and T. Grossman, "Exploring the design space of cognitive engagement techniques with ai-generated code for enhanced learning," in *Proceedings of the 30th international conference on intelligent user interfaces*, 2025, pp. 695–714.

[20] S. Ma, J. Wang, Y. Zhang, X. Ma, and A. Y. Wang, "Dbox: Scaffolding algorithmic programming learning through learner-llm co-decomposition," in *Proceedings of the 2025 CHI Conference on Human Factors in Computing Systems*, 2025, pp. 1–20.

[21] A. authors, "Anonymized tool," 2025. [Online]. Available: https://figshare.com/s/c211fa17f7ef39e34a03

[22] National Institute of Mental Health. (n.d.) Attention-deficit/hyperactivity disorder: What you need to know. [Online]. Available: https://www.nimh.nih.gov/health/publications/attention-deficit-hyperactivity-disorder-what-you-need-to-know

[23] A. P. Fleming and R. J. McMahon, "Developmental context and treatment principles for adhd among college students," *Clinical child and family psychology review*, vol. 15, pp. 303–329, 2012.

[24] B. French, G. Nalbant, H. Wright, K. Sayal, D. Daley, M. J. Groom, S. Cassidy, and C. L. Hall, "The impacts associated with having adhd: an umbrella review," *Frontiers in psychiatry*, vol. 15, p. 1343314, 2024.

[25] D. Adamis, S. West, J. Singh, L. Hanley, I. Coada, G. McCarthy, N. Langan, B. Gavin, and F. McNicholas, "Functional impairment and quality of life in newly diagnosed adults attending a tertiary adhd clinic in ireland," *Irish Journal of Medical Science (1971-)*, vol. 193, no. 5, pp. 2433–2441, 2024.

[26] J. Hernandez-Capistran, G. Alor-Hernandez, L. N. Sanchez-Morales, and I. Machorro-Cano, "A decade of apps for ADHD management: a scoping review," *Behaviour & Information Technology*, pp. 1–28, 2025.

[27] S. Butt, F. E. Hannan, M. Rafiq, I. Hussain, C. N. Faisal, and W. Younas, "Say-it & learn: Interactive application for children with ADHD," in *Cross-Cultural Design. Applications in Health, Learning, Communication, and Creativity: 12th Intl Conference, CCD 2020, Held as Part of the 22nd HCI Intl Conference, HCII 2020, Copenhagen, Denmark, July 19–24, 2020, Proceedings, Part II 22*. Springer, 2020, pp. 213–223.

[28] A. Doulou, A. Drigas, and C. Skianis, "Mobile applications as intervention tools for children with ADHD for a sustainable education." *Technium Sustainability*, vol. 2, no. 4, pp. 44–62, 2022.

[29] A. Doulou, P. Pergantis, A. Drigas, and C. Skianis, "Managing ADHD symptoms in children through the use of various technology-driven serious games: A systematic review," *Multimodal Technologies and Interaction*, vol. 9, no. 1, p. 8, 2025.

[30] L. R. Carvalho, L. M. Haas, G. Zeni, M. M. Victor, S. P. Techele, J. M. Castanho, I. M. Coimbra, A. d. F. de Sousa, N. Ceretta, A. Garrudo *et al.*, "Evaluation of the effectiveness of the focus ADHD app in monitoring adults with attention-deficit/hyperactivity disorder," *European Psychiatry*, vol. 66, no. 1, p. e53, 2023.

[31] L. Powell, J. Parker, V. Harpin *et al.*, "ADHD: is there an app for that? a suitability assessment of apps for the parents of children and young people with ADHD," *JMIR mHealth and uHealth*, vol. 5, no. 10, p. e7941, 2017.

[32] P. Ralph, N. bin Ali, S. Baltes, D. Bianculli, J. Diaz, Y. Dittrich, N. Ernst, M. Felderer, R. Feldt, A. Filieri, B. B. N. de França, C. A. Furia, G. Gay, N. Gold, D. Graziotin, P. He, R. Hoda, N. Juristo, B. Kitchenham, V. Lenarduzzi, J. Martínez, J. Melegati, D. Mendez, T. Menzies, J. Molleri, D. Pfahl, R. Robbes, D. Russo, N. Saarimäki, F. Sarro, D. Taibi, J. Siegmund, D. Spinellis, M. Staron, K. Stol, M.-A. Storey, D. Taibi, D. Tamburri, M. Torchiano, C. Treude, B. Turhan, X. Wang, and S. Vegas, "Empirical standards for software engineering research," 2020. [Online]. Available: https://arxiv.org/abs/2010.03525

[33] S. Santhanam, T. Hecking, A. Schreiber, and S. Wagner, "Bots in software engineering: a systematic mapping study," *PeerJ Computer Science*, vol. 8, p. e866, 2022.

[34] M. Wessel, M. A. Gerosa, and E. Shihab, "Software bots in software engineering: benefits and challenges," in *Proc. of the 19th International Conference on Mining Software Repositories*, ser. MSR '22. New York, NY, USA: Association for Computing Machinery, 2022, p. 724–725. [Online]. Available: https://doi.org/10.1145/3524842.3528533

[35] Y. Gao, Y. Xiong, X. Gao, K. Jia, J. Pan, Y. Bi, J. Sun, and H. Wang, "Retrieval-augmented generation for large language models: A survey."

[36] G. Dong, Y. Zhu, C. Zhang, Z. Wang, J.-R. Wen, and Z. Dou, "Understand what llm needs: Dual preference alignment for retrieval-augmented generation," in *Proceedings of the ACM on Web Conference 2025*, 2025, pp. 4206–4225.
[37] T. Li, X. Hu, and X. Xu, "Design and evaluation of llm-based conversational virtual characters to assist adults with ADHD," in *Intl Conference on Human-Computer Interaction*. Springer, 2025, pp. 360–376.
[38] B. Carik, K. Ping, X. Ding, and E. H. Rho, "Exploring large language models through a neurodivergent lens: Use, challenges, community-driven workarounds, and concerns," *Proceedings of the ACM on Human-Computer Interaction*, vol. 9, no. 1, pp. 1–28, 2025.
[39] S. Baltes and P. Ralph, "Sampling in software engineering research: A critical review and guidelines," *Empirical Software Engineering*, vol. 27, no. 4, p. 94, 2022.
[40] J. S. Molléri, K. Petersen, and E. Mendes, "Survey guidelines in software engineering: An annotated review," in *Proceedings of the 10th ACM/IEEE international symposium on empirical software engineering and measurement*, 2016, pp. 1–6.
[41] D. S. Cruzes and T. Dyba, "Recommended steps for thematic synthesis in software engineering," in *2011 international symposium on empirical software engineering and measurement*. IEEE, 2011, pp. 275–284.
[42] G. Guest, A. Bunce, and L. Johnson, "How many interviews are enough? an experiment with data saturation and variability," *Field methods*, vol. 18, no. 1, pp. 59–82, 2006.